# Bundle adjustment of Hayabusa2's ONC images and controlled color mosaic map of Ryugu

Authors

**Naoyuki Hirata [a], Eri Tatsumi [b,c,d], Mayumi Ichikawa [e], Kazuhiro Honda [e], Sayuri Tanaka [f]**

\* Corresponding Author E-mail address: hirata@tiger.kobe-u.ac.jp

### Authors' affiliation

[a] Graduate School of Science, Kobe University, Kobe, 657-8501, Japan

[b] Instituto de Astrofísica de Canarias (IAC), La Laguna, 38205, Spain

[c] Departamento de Astrofísica, Universidad de La Laguna, La Laguna, 38206, Spain

[d] Department of Earth and Planetary Science, University of Tokyo, Tokyo, 133-0033, Japan

[e] Institute of Space and Astronautical Science, JAXA, Sagamihara, 252-5210, Japan

[f] Japan Manned Space Systems Corporation, Tsukuba, 305-8505, Japan

### Editorial Correspondence to:

Dr. Naoyuki Hirata

Kobe University, Rokkodai 1-1 657-8501

Tel/Fax +81-7-8803-6566

Abstract

JAXA's Hayabusa2 mission successfully returned samples from the asteroid Ryugu in December 2020. It executed two touchdowns to collect the surface and subsurface materials, one close to the crater created by an artificial impactor. The onboard camera system, Optical Navigation Camera (ONC), with two wide-angle cameras and one narrow-angle camera with seven color filters, was crucial for mapping geomorphology and composition such as hydrated minerals during navigation and scientific observation. More than 8,300 images revealed Ryugu's spinning-top shape and boulder-covered surface. However, most high-resolution images captured during descent/touchdown operations lacked precise location data and camera position/orientation information. Image geometry was refined using photogrammetric bundle adjustment. This method enabled the refinement of all high-resolution images captured during descent/touchdown operations. Furthermore, map-projected GeoTIFF images in GIS format containing geographic metadata were created for all ONC images, and these were integrated to construct global and regional mosaic maps. To facilitate scientific research on Ryugu, these refined image geometry information, maps, and mosaics are publicly available via https://doi.org/10.7910/DVN/WW3IH0

# 1. Introduction

Japan Aerospace Exploration Agency's (JAXA) Hayabusa2 mission represents a landmark achievement in asteroid sample-return exploration. After arriving at the near-Earth asteroid (162173) Ryugu on June 27, 2018, Hayabusa2 has conducted numerous proximity operations, including lander deployments, sampling operations, and excavation of surface materials using an artificial impactor. Hayabusa2 deployed small landers, MINERVA-II1A/B (Yoshikawa et al. 2022) and MASCOT (Mimasu et al. 2022), which successfully performed hopping maneuvers and captured high-resolution, close-up imagery and other scientific measurements of the surface. The mission executed its first touchdown (TD1) on February 21, 2019, utilizing a sampling horn and a projectile-driven mechanism to retrieve surface material (Terui et al. 2022). To access pristine subsurface samples, the spacecraft deployed a Small Carry-on Impactor (SCI) on April 5, 2019, creating an artificial crater (Arakawa et al. 2020). This was followed by a second successful touchdown (TD2) on July 11, 2019, in the vicinity of the impact site, ensuring the collection of material shielded from space weathering (Terui et al. 2022). Following its departure from Ryugu in November 2019, Hayabusa2 returned the sample capsule to Earth on December 6, 2020, landing in Woomera, Australia (Nakazawa et al. 2022).

The Optical Navigation Camera (ONC) system (Kameda et al. 2017) onboard the Hayabusa2 spacecraft was vital for both navigation and scientific observation of Ryugu. The ONC system consists of three cameras: a telescopic camera (ONC-T) and two wide-angle cameras (ONC-W1 and ONC-W2). ONC-T is the primary scientific camera, equipped with a filter wheel with seven bandpass filters in the visible and near-infrared range (0.4–0.95 μm, Tatsumi et al. 2019). This multi-band imaging allowed for the capture of detailed color and composition data of the asteroid's surface from various altitudes. ONC-W1 and ONC-W2 were primarily used during decent operations, such as the touchdown sequences, to capture a wide field-of-view (FOV). ONC-W1 has a nadir (downward-facing) view, while ONC-W2 is slant-viewing.

The images captured by the ONC have led to several key scientific findings. The ONC captured more than 8,300 images of Ryugu, revealing its spinning-top shape and rocky surface covered by numerous boulders and pebbles (Watanabe et al. 2019; Sugita et al. 2019). All ONC image data are now available via Data Archives and Transmission System (DARTS)[1] at JAXA and Planetary Data System (PDS)[2] at NASA (Sugita et al. 2022). Stereoscopic images created from ONC-T data allowed for 3D analysis of the asteroid's topography (e.g., Arakawa et al. 2020, Hirata et al. under review). Using the seven color filters, the ONC-T performed multi-band mapping of the surface. This allowed scientists to identify and map the distribution of different surface materials and search for areas rich in hydrated minerals (e.g., Sugita et al. 2019, Tatsumi et al. 2021). Moreover, the ONC is vital for optical navigation, ensuring the spacecraft could accurately approach and rendezvous with the asteroid from millions of kilometers away to complement the radio navigation (Ono et el. 2022). The high-resolution images captured during descents (up to 1 mm/pixel at lowest altitudes) were critical for selecting safe sampling sites and providing geological context for the returned samples (Kikuchi et al. 2020).

---

[1] https://darts.isas.jaxa.jp/missions/hayabusa2/

[2] https://sbn.psi.edu/pds/resource/hayabusa2/

The objective of this study is to refine the geometry information (scene and pose of camera) of all images captured by ONC, using a photogrammetric bundle adjustment on a group of overlapping images. In particular, the geometry information of images captured during descent operations (including MINERVA-II1 and MASCOT deployments, TD1/TD2 touchdown operations, and SCI crater observations) was not created or lacks practical accuracy. The software called StereoPhotoclinometry (SPC) (Gaskell et al. 2008) was used to create a global shape model of Ryugu. As a byproduct of creating the shape model, the precise image geometry information for approximately 7,600 ONC images was obtained (Watanabe et al. 2019). The mosaic maps were created only using those images with the SPC-driven geometry so far (Honda et al. under review). This SPC-driven image geometry information is organized in a data format known as SUMFILE (Gaskell et al. 2008). The image geometry information has been created for images captured from altitudes of 5 km or higher (i.e. global observations such as BOX-C or observations from home position); however, the SUMFILE data for images taken at lower altitudes (such as during TD1 and TD2 or their rehearsals) has not been created. Therefore, many highest-resolution images of Ryugu lack the fundamental data for understanding the imaging area and imaging conditions, and for creating higher-level products such as maps.

In addition to reporting the refined image geometry information after our bundle adjustment, we created (1) Geographic Information Systems (GIS)-formatted maps, GeoTIFF-formatted map-projected files (e.g. Ritter and Ruth, 1997), converted from the ONC images using the above refined image geometry information and the shape model and (2) global and local mosaic maps combined from these maps. GeoTIFF files, a standard Tagged Image File Format (TIFF) file includes embedded geographic metadata as tags within the file's header. This metadata allows the raster image data (pixels) to be accurately geolocated and positioned in the correct real-world location within GIS software, such as QGIS or ArcGIS. A global mosaic and various regional/local mosaics were generated by integrating the aforementioned map projection products. Refined image geometry information, maps, and mosaics are publicly available and freely accessible, and contribute to fully organizing Hayabusa2's imagery into a library. We expect that scientists and students will be able to access image data more easily and contribute to generating science outcomes using those products.

2. Data

Primary mission phase, operations and image resolution acquired from each phase are summarized in Table 1 (see also Tsuda et al. 2022, Watanabe et al. 2025, and reference therein) and Figure 1. Global resolution map of all ONC-T images are shown in Figure 2 (this is one of the primary output of this research). Due to the extremely weak gravitational field of asteroid Ryugu, the Hayabusa2 spacecraft did not enter a closed orbit. Instead, the spacecraft maintained a controlled hovering state at a nominal altitude of approximately 20 km. This primary operational base was designated the "home position" (or BOX-A). Images acquired by the ONC-T from this altitude provided a spatial resolution of 2.2 m/px. More than 7,000 ONC images are acquired from the home position. Throughout the mission, the spacecraft executed various proximity operations from the home position, including descents to altitudes below 6 km for high-resolution imaging, the deployment of rovers and a lander, two touchdown maneuvers for sample collection, and the artificial crater searching observations.

Hayabusa2 was designed to remain positioned along the Earth-asteroid line. Given that Ryugu's rotational axis is nearly perpendicular to its orbital plane (with an inclination of 171.6°, Watanabe et al. 2019), the spacecraft was constrained to the equatorial plane. While this geometry facilitated observations of the equatorial region across diverse emission angles, it was poorly suited for polar observations, as these regions could only be viewed at high, oblique angles. To mitigate this, the spacecraft was transitioned from BOX-A toward the poles to enable nadir-looking observations, an operation known as BOX-B. These maneuvers were performed for the south pole on August 24, 2018 (2.6 m/px resolution), and for the north pole on January 24, 2019 (2.1 m/px). Furthermore, the illumination conditions, including phase angles, varied significantly between operational phases. In 2018 (prior to solar conjunction), shadows of positive relief features, such as boulders, were cast toward the east. Conversely, in 2019 (following solar conjunction), these shadows fell toward the west. Consequently, despite covering the same terrain, images from 2018 and 2019 often exhibit markedly different visual characteristics due to these opposing lighting geometries.

High-resolution regional observations were performed during nine operations: BOX-C 1–7, Medium Altitude Observations 1–2, and Crater Search Operations CRA1 and CRA2. During these phases, the spacecraft's altitude ranged from approximately 8 km to 1.7 km. These operations facilitated wide-area imaging with varying spatial resolutions (Figure 2): 0.87 m/px for regions north of 70°S, 0.66 m/px south of 70°N, 0.55 m/px between 30°N and 50°S, and 0.28 m/px within the 13°N–27°S latitudinal band. Additionally, to assess surface morphological changes induced by the artificial impact experiment, regional imaging was conducted at a higher resolution of 0.18 m/px, covering the longitudinal range of 260°–340°E.

Local observations at much higher resolutions were acquired during 11 distinct descent operations: gravity-measurement, MINERVA-II1 and MASCOT deployments, and the TD1 and TD2 operations and their associated rehearsals. The spacecraft repeatedly visited the TD1 site three times and the TD2 site five times, capturing the highest-resolution images of Ryugu, reaching approximately 0.35 mm/px.

Table 1 Minimum altitude and best image resolution during image acquisition in each primary mission phase from https://data.darts.isas.jaxa.jp/pub/pds4/data/hyb2/hyb2/document/hyb2_mission_event_timeline.pdf

| Date (YYYYMMDD) | Mission phase | Min. Altitude (m) | Best image resolution (m/px) |
|---|---|---|---|
| 20180627 | Arrival at Ryugu (home position) | 20,000 | 2.2 |
| 20180720 | BOX-C1 operation | 6,300 | 0.68 |
| 20180801 | Medium altitude observations 1 | 5,100 | 0.55 |
| 20180806 | Gravity measurement descet operation | 860 | 0.092 |
| 20180824 | BOX-B1 operation to the south pole | 24,000 | 2.6 |
| 20180831 | BOX-B2 operation to the dusk direction | 21,000 | 2.1 |
| 20180911 | Rehearsal 1 for Touchdown 1 (TD1-R1) | 2,900 | 0.28 |
| 20180921 | MINERVA-II1 deployment operation | 57 | 0.0062 |
| 20181003 | MASCOT deployment operation | 37 | 0.0051 |
| 20181003 | Medium altitude observations 2 | 2,600 | 0.28 |
| 20181015 | Rehearsal 1A for Touchdown 1 (TD1-R1-A) | 40 | 0.0043 |
| 20181025 | Rehearsal 3 for Touchdown 1 (TD1-R3) | 40 | 0.0043 |
| 20181030 | BOX-C2 operation | 5,600 | 0.60 |
| 20181101 | BOX-C3 operation | 2,200 | 0.24 |
| 20190108 | BOX-B3 operation to the subsolar direction | 20,000 | 2.0 |
| 20190124 | BOX-B4 operation to the north pole | 20,000 | 2.1 |
| 20190221 | Touchdown 1 (TD1) | 1.2 | 0.0015 |
| 20190228 | BOX-C4 operation | 5,700 | 0.61 |
| 20190308 | Descent observation (DO-S01) for TD2 | 21 | 0.027 |
| 20190321 | Crater search operation before SCI (CRA1) | 1,700 | 0.18 |
| 20190425 | Crater search operation after SCI (CRA2) | 1,700 | 0.18 |
| 20190516 | Low altitude observation and target marker separation operation (PPTD-TM1) | 49 | 0.062 |
| 20190530 | Low altitude observation and target marker separation operation (PPTD-TM1A) | 30 | 0.0034 |
| 20190613 | Low descent observation operation (PPTD-TM1B) | 30 | 0.0032 |
| 20190711 | Touchdown 2 (TD2) | 1.3 | 0.0017 |
| 20190725 | BOX-C5 operation | 5,000 | 0.54 |
| 20190813 | BOX-B5 operation to the dusk direction | 20,000 | 2.2 |
| 20190823 | BOX-B6 operation to the dawn direction | 20,000 | 2.2 |
| 20190923 | BOX-B7 operation from dusk to subsolar direction | 20,000 | 2.2 |
| 20191005 | BOX-C6 operation | 7,500 | 0.80 |
| 20191024 | BOX-C7 operation | 4,400 | 0.47 |
| 20191113 | Departure | 20,000 | 2.2 |

## 3. Method

We performed a rigorous photogrammetric bundle adjustment on overlapping image sets to determine the precise geometric parameters for ONC images acquired during the descent operations. A significant challenge in this process was identifying the exact imaging areas, many of which were initially unknown. This difficulty arises from several factors: (1) the substantial resolution disparity, often exceeding an order of magnitude, precluded automated methods such as template matching and hindered visual identification; (2) the LIDAR-based kernels were insufficient due to the lack of precise camera orientation; and (3) the wide-angle (ONC-W) and telescopic (ONC-T) cameras were not operated simultaneously, causing ONC-T images to become spatially isolated. Consequently, identifying these areas requires meticulous manual inspection, inferring positions from the temporal sequence of preceding and succeeding frames. This manual verification constituted the most time-consuming phase of the workflow. Nevertheless, it was a critical step, as it defined the feature pairings essential for the subsequent generation of control point networks.

Bundle adjustment is an optimization technique in image processing and computer vision that simultaneously refines the 3D coordinates of points in a scene and the pose (position and orientation) of the cameras that captured the images to achieve globally consistent and accurate 3D reconstructions. The fundamental idea of bundle adjustment is to find the set of camera and 3D point parameters that minimizes the reprojection error. The reprojection error is the pixel distance between the observed 2D location of a feature point in an image and the projected 2D location of its corresponding 3D point based on the current estimates of image geometry information. In this study, we developed a new tool to iteratively solve this process. This tool utilizes the open-source Python libraries SciPy (Virtanen et al. 2020) and Numpy (Harris et al. 2020) for scientific computing, along with OpenCV (Bradski and Kaehler, 2008) and Open3D (Zhou et al. 2018) for computer vision and image processing, solving the process using the Levenberg-Marquardt optimization algorithm.

In detail, we first searched for common feature points within a group of overlapping images based on visual characteristics to create a set of control points; by connecting images with high-precision geometry information with images with low-precision geometry information within the group, we refined the latter geometry information. We then selected images with high-precision geometry information as reference images, by comparing them with the asteroid's edge and shape model. Feature detection and extraction were performed using either the template matching algorithm from the OpenCV library or a fully manual process involving human observation and manual selection. The control points are distributed as widely as possible across the entire overlapping area of the images, ensuring that each image contains at least 12 valid points. In particular, images captured immediately after touchdown show scattered ejecta particles, moving boulders, and the large shadow of the spacecraft, which can cause false detections during automated searches for the same location; therefore, for these images, the generated control points were carefully verified by human eye to eliminate false detections. As a result, more than 10,000 control points were generated to refine the image geometry information. Using these control points, the geometry information of images acquired during descent operations is estimated by referencing the shape model and/or ones pre-estimated by SPC. Then, the 3D coordinates of the control points are constrained to align with the shape model. Although the camera position estimates from LIDAR are available, they were not utilized in this study. We used the ONC L2b images, which have undergone flat-field correction but have not yet undergone camera distortion correction, for the estimation of image geometry information. Although L2d images, undistorted and photometrically corrected images, are available, we do not use L2d images because their edge pixels were cropped. To maximize the usable imaged area, we applied internal distortion correction to the image coordinates of the L2b images in the tool. Imaging through the color filters of ONC-T involves slightly different focal lengths, but here we assumed a fixed focal length of 120.5 mm as the v-band image, which is absorbed by the distance from Ryugu's surface for images captured by other bands than the v-band.

      Using the shape model and updated image geometry information we converted the ONC images into to GIS-formatted maps. The maps are created with a resolution 1.2 times higher than that of the original image. A latitude-longitude grid is constructed on the surface of the 3D shape model and projected and ray-casted onto the image plane using the refined image geometry information to obtain the image coordinates for each grid point. The values at each grid point are estimated from the surrounding pixel values using bicubic interpolation. During this process, occlusion handling was implemented to prevent the projection of invisible regions (areas where the line of sight from the camera's position is obstructed by other parts of the asteroid's polygonal mesh). No further image processing that moves pixels, such as homography or affine transformations, was performed. Additionally, we created spatial resolution, emission angle, and solar incidence angle maps corresponding each L2b/L2d image map using the shape model and the image geometry information. The resulting map data are also utilized to evaluate the accuracy of the bundle adjustment and updated image geometry information; if it is sufficiently accurate, the maps should match with each other without

further image processing.

## 4. Result

Consequently, the bundle adjustment performed in this study significantly refined the geometric information for 994 ONC images. As a representative example, we present regional mosaic maps of the western hemisphere acquired during the artificial crater search operations (CRA1 and CRA2). CRA1 and CRA2 observed almost the same region, but CRA1 was conducted on 20190321 (YYYYMMDD) which is before the Small Carry-on Impactor (SCI) experiment while CRA2 was conducted on 20190425 after the experiment. Previously, CRA1 and CRA2 mosaic maps exhibited misalignments; furthermore, inconsistencies persisted between maps produced using different color filters, even for observations conducted on the same day. In the current version, these mosaic maps are well-aligned, facilitating high-precision surface comparisons of Ryugu before and after the SCI experiment. This improved geometric consistency enables pixel-to-pixel change detection of the ground surface with unprecedented clarity. Additionally, the geometry information for all images captured during the 11 descent operations, including the geometry of touchdown images characterized by ejecta scattering, were optimized. The resulting maps demonstrate that the background terrain remains fixed while ejecta particles disperse, facilitating high-resolution analysis of particle distribution and trajectory with greater precision than previously possible.

A total of 8,357 ONC map-projected images, including those processed with the SPC-based and our bundle adjustment metadata, are available at the following link (see Summary). To ensure usability for a broad research community, including non-experts in image processing, the dataset provides geometry information in SUMFILE format, GeoTIFF map images, and summary text files containing corner coordinates, resolution, and geographic bounds.

For approximately 100 images, geometric estimation remained unfeasible due to technical difficulties in identifying sufficient control points for the bundle adjustment. These cases primarily include the images captured at the moment of touchdowns, where the frames are dominated by scattered ejecta and the shadow of the Hayabusa2 spacecraft, the images taken from large distances where the asteroid appears too small for finding control points, or the images which capture only a small portion of the asteroid's relative to the total FOV.

Alignment of images captured during the descent operations is affected by parallax-induced registration errors, stemming from the size of polygons and accuracy of the shape model. For high-resolution descent images less than 0.1 m/px (~60 pixels/degree), this insufficient accuracy or resolution of the shape model prevents precise orthorectification. Consequently, small-scale topographic features like boulders, if absent from the shape model, are incorrectly projected and appear elongated. Although a higher-fidelity shape model could mitigate these errors, its development was beyond the scope of this. The impact of parallax is most pronounced in the wide-angle ONC-W1/W2 images, where large emission angles close to the edge of FOV amplify projection errors. Conversely, global and some regional observation data (BOX-C, medium altitude observations, or gravity measurements) show negligible misalignment; for these data, the shape model resolution is sufficiently high to achieve high accuracy orthorectification. Similarly, ONC-T images taken during descending remain largely unaffected by parallax shifts, as its near-nadir viewing geometry and narrow FOV angle ensures a near-orthographic projection.

We combined these maps to create global and regional mosaic maps. As an example, the global mosaic map (Figure 2) was composited from the medium altitude observations 1 on 20180801 (Table 1), BOX-C7 operation on 20191025, BOX-C4 operation on 20190228 and 20190301, BOX-B4 operation to the north pole on 20190124, and BOX-B1 operation to the south pole on 20180824. To generate the global map, several preprocessing steps were applied prior to mosaicking: pixels with emission angles exceeding 65 degrees were excluded, non-overlapping pixels across the seven-band synchronous images were removed, and the brightness and contrast of L2d images were normalized to match the seven-band data from 20180801. In addition, we created some regional mosaics for the western hemisphere before and after the SCI experiment created from the four-band observations during crater searching operations on 20190321 and 20190425, the equatorial region created from the seven-band observations during the medium altitude observation 2 in 20181003-20181004, and the northern and southern hemisphere in polar azimuthal equidistant projection during the BOX-B operations to the north and south poles on 20180824 and 20190124 (Figure 3). Furthermore, we created local mosaics from images obtained during the 11 descent operations. These data are grouped into four distinct regions (TD1, TD2, MINERVA, and MASCOT), where local mosaics were constructed to define the spatial coverage of each image. The mosaic map around TD1 site were created from the three descent observations (20181015, 20181025, and 20190221-20190222), and that around TD2 site with the five descent observations (20190308, 20190516, 20190530, 20190613, and 20190711). These mosaic maps are provided as baseline examples. We anticipate that researchers, students, and science communicators will utilize tools such as the GDAL libraries and command-line utilities to generate custom mosaic maps tailored to their specific scientific objectives.

Table 2 List of mosaic maps

| mosaic name | Resolution (pixel per degree) | description |
| --- | --- | --- |
| ryugu_global.tif | 10 | global mosaic |
| ryugu_northpole.tif | 10 | from BOX-B4 operation to the north pole on 20190124 in polar projection |
| ryugu_southpole.tif | 10 | from BOX-B1 operation to the south pole on 20180824 in polar projection |
| ryugu_midalt.tif | 40 | from medium altitude observations 2 on 20181003 |
| ryugu_20190321.tif | 60 | from artificial crater search operation (CRA1) on 20190321 (before the SCI experiment) |
| ryugu_20190425.tif | 60 | from artificial crater search operation (CRA2) on 20190425 (after the SCI experiment) |
| ryugu_gravity.tif | 80 | from operation to measure the gravity of Ryugu on 20180806 |
| ryugu_minerva*.tif | 120-1450 | from MINERVA-II1 deployment operation on 20180921 |
| ryugu_mascot*.tif | 140-1750 | from MASCOT deployment operation on 20181003 |
| ryugu_td1e*.tif | 160-5800 | ONC-W mosaics from the three descent observations for TD1 (20181015, 20181025, and 20190221) |
| ryugu_td1h*.tif | 40-2100 | ONC-T mosaics from the three descent observations for TD1 (20181015, 20181025, and 20190221) |
| ryugu_td2f*.tif | 160-5200 | ONC-W mosaics from the five descent observations for TD2 (20190308, 20190516, 20190530, 20190613, and 20190711) |
| ryugu_td2g*.tif | 30-3000 | ONC-T mosaics from the five descent observations for TD2 (20190308, 20190516, 20190530, 20190613, and 20190711) |

## 5. Summary

Our bundle adjustment significantly refined image geometric information for 994 ONC images, enabling creation of global and regional mosaic maps of Ryugu. This improved geometric consistency, for example, facilitates high-precision surface comparisons of Ryugu before and after the SCI experiment, allowing pixel-to-pixel change detection of the ground surface with unprecedented clarity. Geometric information for all images from 11 descent operations, including touchdowns, was optimized, demonstrating fixed background terrain and dispersing ejecta particles for high-resolution analysis. A total of 8,357 ONC images with the SPC-based and our updated geometric metadata are available, featuring geometry information in SUMFILE format, GeoTIFF map-projected images, and summary text files with corner coordinates, resolution, and geographic bounds for broad research use. Comprehensive guidelines for data processing and integration using GDAL libraries are included. Alignment of descent images (especially by ONC-W1 and W2) is affected by parallax-induced registration errors from the insufficient resolution of the shape model, which can cause small-scale topographic features like boulders to be incorrectly projected and appear elongated in high-resolution images. However, the global observation data and ONC-T images taken during descents are largely unaffected by parallax shifts because of sufficient shape model resolution or near-nadir viewing geometry. Created mosaic maps include the global map composited from various observations and the regional maps for areas like the western hemisphere before and after the SCI experiment, the equatorial regional map, and the northern/southern hemisphere maps in polar azimuthal equidistant projection. Local mosaics were also created from images obtained during the 11 descent operations for four distinct regions (TD1, TD2, MINERVA, and MASCOT), defining the spatial coverage of each image. Users are expected to freely create original mosaic maps using these data and tools like GDAL libraries.

The link (https://doi.org/10.7910/DVN/WW3IH0) contains all GeoTIFF map files created from ONC images (L2b and L2d) and mosaics combined from the maps, along with camera image geometry information in SUMFILE format, licensed under CC BY 4.0. The spatial relationships and general information of the mosaics are shown in "./ReadMeFirst.pptx". Comprehensive guidelines for download URL, data processing, and integration using GDAL libraries are included in "./ReadMeSecond.txt". Specific geometry information of each ONC image, such as minimum/maximum values of resolution/latitude/longitude, the distance from Ryugu, the latitude/longitude on Ryugu's coordinate system at the image corners and image center, is summarized in "./onc_camerainfo.txt". The directory of "./mosaic/" contains all mosaics created by this research. The directory of "./map" contains map files corresponding to each ONC L2b/L2d image, including north and south polar azimuthal equidistant projection maps, simple cylindrical projection maps, and their emission angle, solar incidence angle, phase angle, and resolution maps. We would appreciate it if you could cite this paper when using them.

Acknowledgments

We would like to thank all members of Hayabusa2 mission team for their support of the data acquisition. This study was partly supported by the JSPS Grants-in-Aid for Scientific Research (Nos. 20K14538 and 20H04614) and Hyogo Science and Technology Association. This study was supported by the JAXA Hayabusa2# International Visibility Enhancement Project.

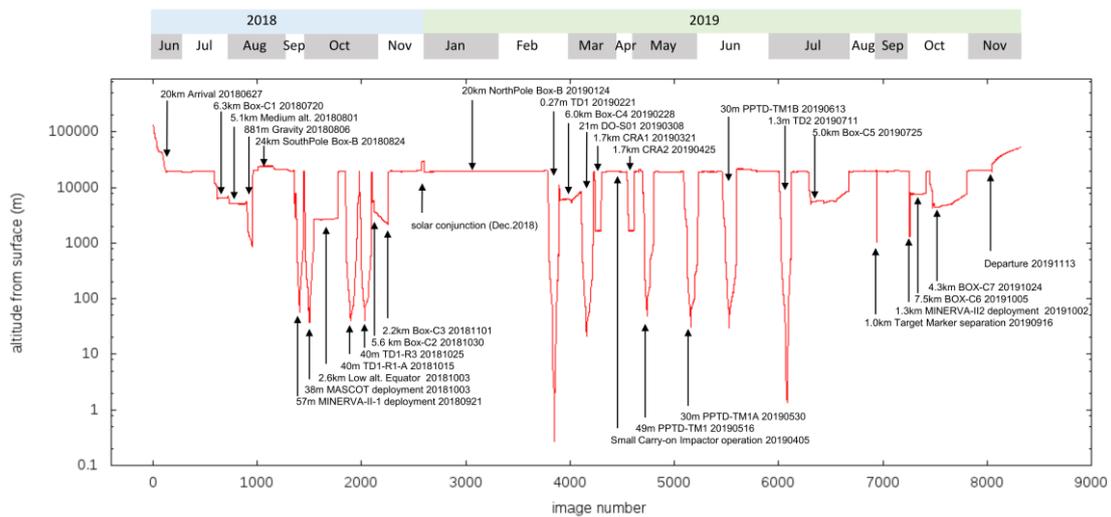

Figure 1 Hayabusa2's altitude from the surface at the time of image capture. The horizontal axis shows the cumulative number of ONC images capturing Ryugu.

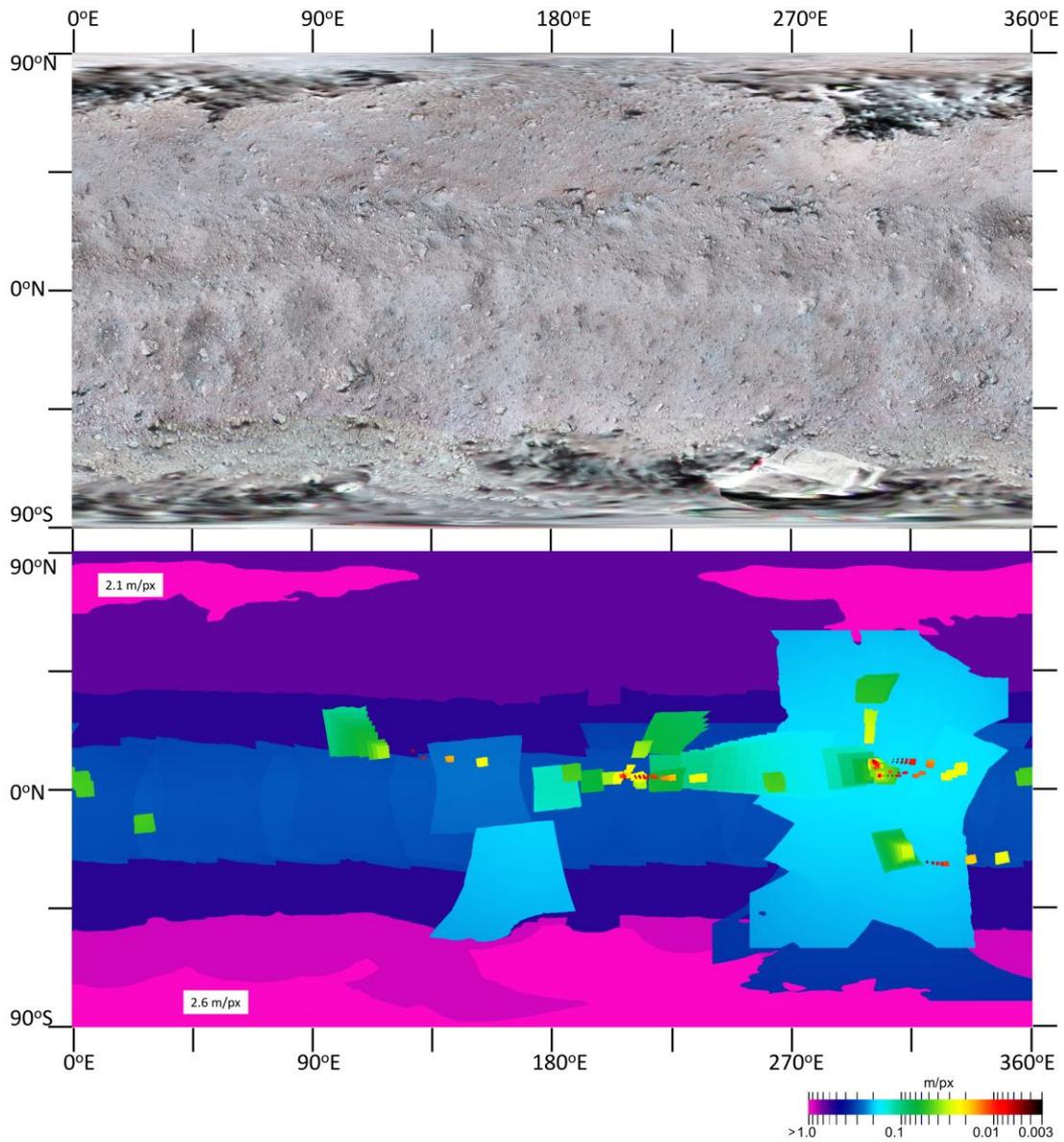

**Figure 2** (Top) Global color mosaic of Ryugu, with the blue, green, and red channels assigned to the ul-band, v-band, and p-band, respectively. (Bottom) Global resolution map of ONC-T images.

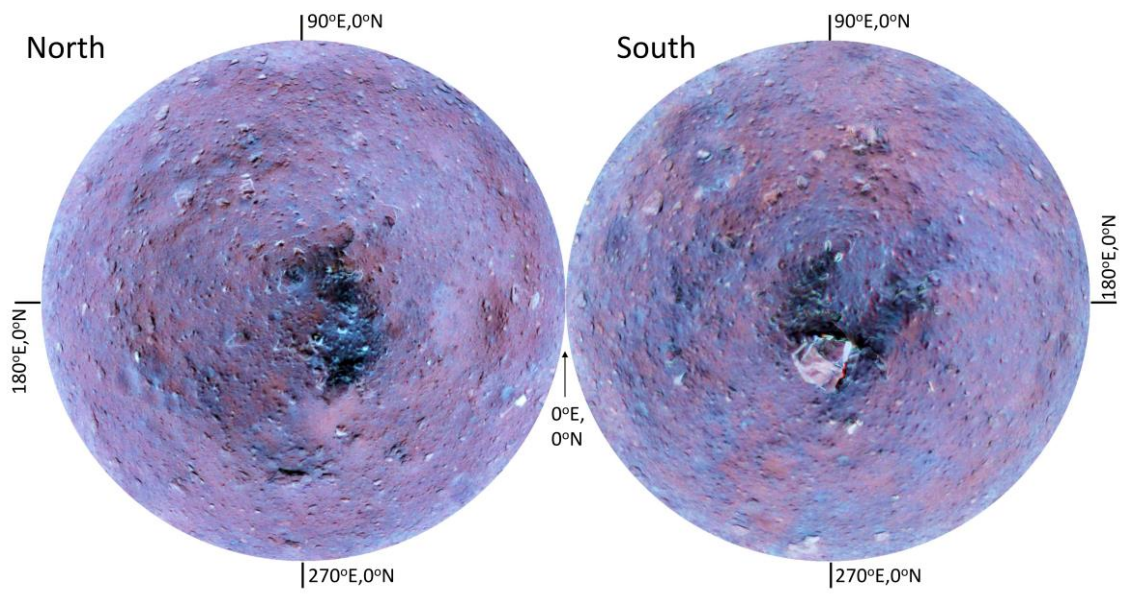

**Figure 3** (left) North and (right) south polar azimuthal equidistant projection enhanced color-composite maps, , with the blue, green, and red channels assigned to the ul-band, v-band, and p-band. The circumference is the equator.

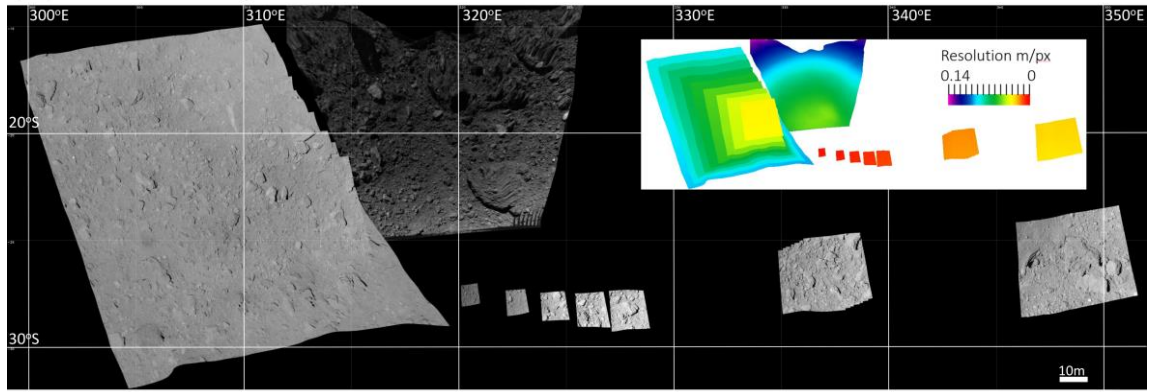

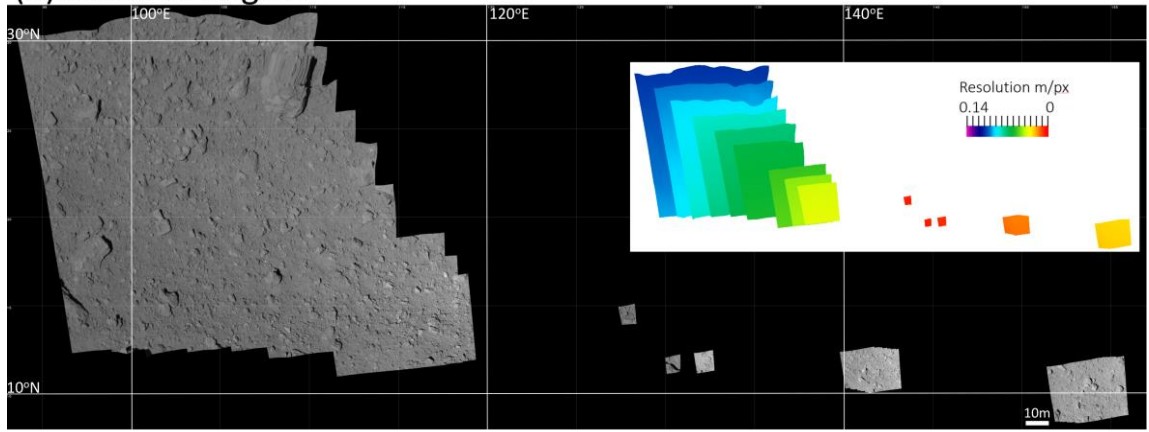

Figure 4 Simple cylindrical projection maps of ONC-T and ONC-W2 images captured during the decent operations for MASCOT (a) and MINERVA (b) deployments.

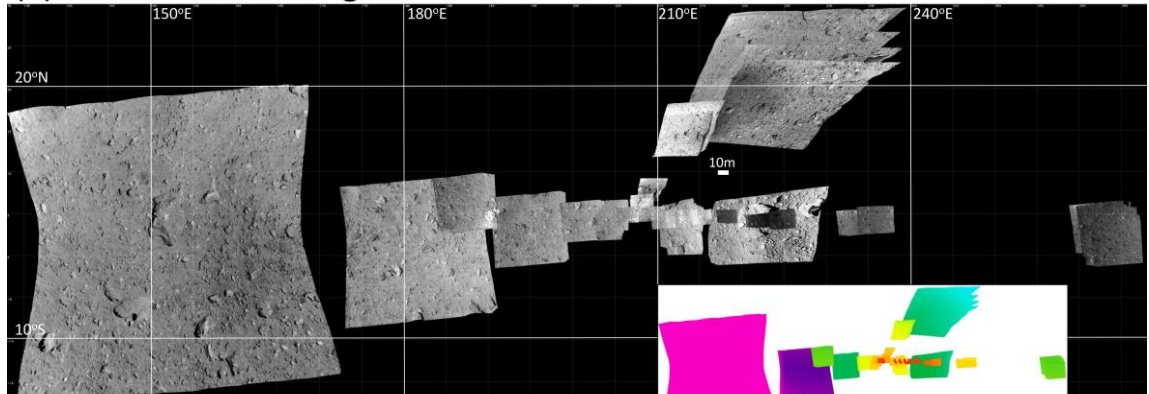

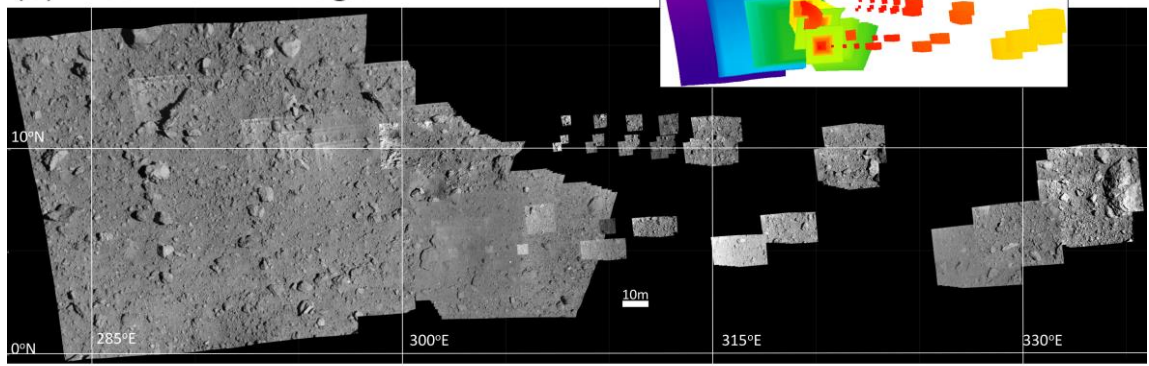

**Figure 5.** Simple cylindrical projection maps of ONC-T images captured during the decent operations for TD1 (a) and TD2 (b).

References

Arakawa, M. et al. (2020) An artificial impact on the asteroid (162173) Ryugu formed a crater in the gravity-dominated regime. Science 368, 67-71.

Bradski, G., and Kaehler, A. (2008). Learning OpenCV: Computer Vision with the OpenCV Library. O'Reilly Media, Inc.

Gaskell, R. W., O. S. Barnouin-Jha, D. J. Scheeres, A. S. Konopliv, T. Mukai, S. Abe, J. Saito, M. Ishiguro, T. Kubota, T. Hashimoto, J. Kawaguchi, M. Yoshikawa, K. Shirakawa, T. Kominato, N. Hirata, H. Demura (2008). Characterizing and navigating small bodies with imaging data. Meteorit. Planet. Sci. 43, 1049–1061.

Harris, C. R., Millman, K. J., van der Walt, S. J., Gommers, R., Virtanen, P., Cournapeau, D., et al. (2020). Array programming with NumPy. Nature, 585(7825), 357-362.

Hirata, N. et al. High-Resolution Digital Terrain Models of Ryugu by the Hayabusa2 Mission, Icarus, in revision

Honda, K., E. Tatsumi, Y. Yokota, D. Shoji, S. Murakami, M. Ichikawa, R. Miyazaki, T.

Kouyama, S. Sugita, R. Honda, T. Morota, S. Kameda, H. Sawada, M. Yamada, H. Suzuki, Y. Cho, N. Sakatani, M. Matsuoka, M. Hayakawa, K. Yumoto, C. Honda, N. Hirata, K. Ogawa, K. Yoshioka, D. Domingue, H. Sato. Mosaic Terrain Maps of (162173) Ryugu. Progress in Earth and Planetary Science, under review.

Kameda, S., H. Suzuki, T. Takamatsu, Y. Cho, T. Yasuda, M. Yamada, H. Sawada, R. Honda, T. Morota, C. Honda, M. Sato, Y. Okumura, K. Shibasaki, S. Ikezawa, S. Sugita (2017). Preflight calibration test results for optical navigation camera telescope (ONC-T) onboard the Hayabusa2 spacecraft. Space Sci. Rev. 208, 17–31.

Kikuchi, S. et al. (2020) Hayabusa2 Landing Site Selection: Surface Topography of Ryugu and Touchdown Safety. Space Sci Rev 216, 116.

Mimasu, Y., K. Yoshikawa, G. Ono, N. Ogawa, F. Terui, Y. Takei, T. Saiki, T. Ho, A. Moussi, Y. Tsuda (2022). MASCOT lander release operation, in Hayabusa2 Asteroid Sample Return Mission, pp.229-240.

Nakazawa, S., K. Kawahara, T. Yamada, N. Fujita, T. Ishimaru, A. Miura, T. Ito, D. Hayashi, K. Fujita, H. Tanno, H. Sawada, S. Tanaka, N. Kobiki, T. Saiki, Y. Tsuda (2022). Hayabusa2 reentry and recovery operations of the sample return capsule, in Hayabusa2 Asteroid Sample Return Mission, Technological Innovation and Advances, pp. 95-111.

Ono, G., N. Ogawa, H. Takeuchi, H. Ikeda, Y. Takei, F. Terui, Y. Mimasu, K. Yoshikawa, T. Saiki, Y. Tsuda (2022). Controlled descent of Hayabusa2 to Ryugu, in Hayabusa2 Asteroid Sample Return Mission, pp. 177-187.

Ritter, N., and Ruth, M. (1997). The GeoTiff data interchange standard for raster geographic images. International Journal of Remote Sensing, 18(7), 1637–1647. https://doi.org/10.1080/014311697218340

Sugita, S. et al. (2019) The geomorphology, color, and thermal properties of Ryugu: Implications for parent-body processes. Science 364, eaaw0422.

Sugita, S., R. Honda, T. Morota, S. Kameda, H. Sawada, Y. Yokota, M. Yamada, T. Kouyama, E. Tatsumi, H. Suzuki, Y. Cho, N. Sakatani, M. Matsuoka, M. Hayakawa, K. Yumoto, C. Honda, K. Ogawa, K. Yoshioka, S. Murakami, Y. Yamamoto, M. K. Crombie (2022). Hayabusa2 ONC Bundle, urn:jaxa:darts:hyb2_onc, JAXA Data Archives and Transmission System, https://doi.org/10.17597/isas.darts/hyb2-00200.

Tatsumi, E. et al. (2019) Updated inflight calibration of Hayabusa2's optical navigation camera (ONC) for scientific observations during the cruise phase. Icarus 325, 153-195.

Tatsumi, E., Sakatani, N., Riu, L. et al. Spectrally blue hydrated parent body of asteroid

(162173) Ryugu. Nat Commun 12, 5837 (2021). https://doi.org/10.1038/s41467-021-26071-8

Terui, F., S. Kikuchi, Y. Takei, Y. Mimasu, H. Sawada, T. Saiki, Y. Tsuda (2022). Touchdown operation planning, design, and results, in Hayabusa2 Asteroid Sample Return Mission, pp. 259-289.

Tsuda, Y., S. Nakazawa, M. Yoshikawa, T. Saiki, F. Terui, M. Arakawa, M. Abe, K. Kitazato, S. Sugita, S. Tachibana, N. Namiki, S. Tanaka, T. Okada, H. Ikeda, S. Watanabe (2022). Mission objectives, planning, and achievements of Hayabusa2, in Hayabusa2 Asteroid Sample Return Mission, pp. 5-23.

Virtanen, P., Gommers, R., Oliphant, T.E. et al. (2020). SciPy 1.0: fundamental algorithms for scientific computing in Python. Nature Methods 17, 261–272. https://doi.org/10.1038/s41592-019-0686-2

Watanabe, S. et al. (2019) Hayabusa2 arrives at the carbonaceous asteroid 162173 Ryugu—A spinning top–shaped rubble pile. Science 364, 268–272.

Watanabe, S. et al. (2025) Hayabusa2 Mission Bundle v4.0, urn:jaxa:darts:hyb2::4.0, JAXA Data Archives and Transmission System, https://doi.org/10.17597/isas.darts/hyb2-00100

Yoshikawa, K., S. V. Wal, Y. Mimasu, N. Ogawa, G. Ono, F. Terui, T. Saiki, Y. Tsuda (2022). MINERVA-II-1A/B Asteroid Rovers Deployment and Landing, in Hayabusa2 Asteroid Sample Return Mission, pp. 209-228.

Zhou, Q. Y., Park, J., & Koltun, V. (2018). Open3D: A modern library for 3D data processing. arXiv preprint arXiv:1801.09847.